\documentclass[twocolumn,showpacs,aps,pre]{revtex4-1}

\usepackage{amssymb,amsmath}
\usepackage[dvips]{graphicx}

\begin{document}

\title{Generalized contact process with two symmetric absorbing states in two dimensions}

\author{Man Young Lee}
\author{Thomas Vojta}
\affiliation{Department of Physics, Missouri University of Science and Technology, Rolla, MO 65409, USA}

\begin{abstract}
We explore the two-dimensional generalized contact process with two absorbing states
by means of large-scale Monte-Carlo simulations. In part of the phase diagram,
an infinitesimal creation rate of active sites between inactive domains is sufficient
to take the system from the inactive phase to the active phase.
The system therefore displays two different nonequilibrium phase transitions. The critical
behavior of the generic transition is compatible with the generalized voter (GV)
universality class, implying that the symmetry-breaking and absorbing transitions coincide.
In contrast, the transition at zero domain-boundary activation rate
is not critical.
\end{abstract}

\date{\today}

\maketitle

%%%%%%%%%%%%%%%%%%%%%%%%%%%%%%%%%%%%%%%%%%%%%%%%%%%%%%%%%%%%%%%%%%%%%%%%%%%%%%%%%
% Main text starts here
%%%%%%%%%%%%%%%%%%%%%%%%%%%%%%%%%%%%%%%%%%%%%%%%%%%%%%%%%%%%%%%%%%%%%%%%%%%%%%%%%
\section{Introduction}
%%%%%%%%%%%%%%%%%%%%%%%%%%%%%%%%%%%%%%%%%%%%%%%%%%%%%%%%%%%%%%%%%%%%%%%%%%%%%%%%%

Phase transitions between different nonequilibrium steady states are a topic of great
current interest in statistical physics. These transitions display large-scale
fluctuations and collective behavior over large distances and long times just as
equilibrium phase transition. They occur, for example, in surface growth, granular flow,
chemical reactions, population dynamics, and even in traffic jams
\cite{ZhdanovKasemo94,SchmittmannZia95,MarroDickman99,Hinrichsen00,Odor04,Luebeck04,TauberHowardVollmayrLee05}.

The so-called absorbing state transitions are a particularly well-studied type of
nonequilibrium phase transitions. They separate fluctuating (active) steady states from absorbing
(inactive) states where fluctuations stop completely. Generically, absorbing state transitions
are in the directed percolation (DP) \cite{GrassbergerdelaTorre79} universality class;
Janssen and Grassberger \cite{Janssen81,Grassberger82} conjectured
that all absorbing state transitions with a scalar order parameter and short-range interactions
belong to this class as long as there are no extra symmetries or conservation laws.
This conjecture has been confirmed in countless theoretical and computer simulation
studies. Experimental verifications were found in ferrofluidic spikes
\cite{RuppRichterRehberg03} and in the transition between two
turbulent states in a liquid crystal \cite{TKCS07}.

In recent years, significant attention has focused on absorbing state transitions
in universality classes different from DP that can occur if the system features
additional symmetries or conservation laws. In 1997, Hinrichsen \cite{Hinrichsen97}
suggested several nonequilibrium stochastic lattice models with $n\ge 2$ absorbing states.
In the case of two symmetric absorbing states ($n=2$), he found the critical exponents
to be different from the DP values. The corresponding universality class has been given
several different names in the literature such as  the $Z_2$-symmetric directed percolation
class (DP2) or the directed Ising (DI) class. If the symmetry
between the two absorbing states is broken, the critical behavior reverts back to DP.

Recently, we revisited \cite{LeeVojta10} one of the stochastic lattice models introduced in Ref.\
\cite{Hinrichsen97}, \emph{viz.}, the generalized contact process with two absorbing states in one
space dimension. By employing large-scale Monte-Carlo simulations, we found a rich phase diagram
featuring two different nonequilibrium phase transitions separated by a special point that shares
some characteristics with a multicritical point. The generic transition occurs at nonzero values
of the infection, healing and domain-boundary activation rates. It belongs to the above-mentioned
DP2 or DI universality class
which in one dimension coincides \cite{Hinrichsen00} with the
parity-conserving (PC) class \cite{GrassbergerKrauseTwer84} (occurring, e.g., in the
branching-annihilating random walk with an even number of offspring (BARWE)
\cite{ZhongAvraham95}).  In addition, we found an unusual line of phase transitions at zero
domain-boundary activation rate which turned out to be non-critical.

Here, we consider the generalized contact process with two symmetric absorbing states in \emph{two}
space dimensions. The purpose of this paper is twofold. First, we wish to investigate whether the
two-dimensional generalized contact process also displays the above-mentioned rich phase diagram
having two nonequilibrium phase transitions. Second, we wish to study the critical behavior of
these transitions and their universality. According to a conjecture by Dornic \emph{et al.}
\cite{DCCH01}, transitions with $Z_2$ symmetry and no bulk fluctuations (i.e., transitions with
two symmetric absorbing states) should be in the generalized voter (GV)
universality class for which the upper critical dimension is exactly two. Alternatively, the transition
could split into a symmetry-breaking Ising transition and a DP transition
\cite{DrozFerreiraLipowski03,ACDM05}. To address these questions, we perform large-scale Monte-Carlo
simulations.

Our paper is organized as follows. We introduce the generalized contact process with
several absorbing states in Sec.\ \ref{sec:processes}. Sec.\ \ref{sec:simulations} is devoted
to the results and interpretation of our Monte-Carlo simulations. We conclude in Sec.\
\ref{sec:conclusions}.

%%%%%%%%%%%%%%%%%%%%%%%%%%%%%%%%%%%%%%%%%%%%%%%%%%%%%%%%%%%%%%%%%%%%%%%%%%%%%%%%%
\section{Generalized contact process with several absorbing states}
\label{sec:processes}
%%%%%%%%%%%%%%%%%%%%%%%%%%%%%%%%%%%%%%%%%%%%%%%%%%%%%%%%%%%%%%%%%%%%%%%%%%%%%%%%%

We first define the simple contact process \cite{HarrisTE74}, one of the  prototypical models
in the DP universality class. Each site $\mathbf{r}$ of a $d$-dimensional hypercubic lattice
can be in one of two states, either A, the active (infected) state or I, the inactive (healthy)
state. During the time evolution of the contact process, active sites infect their nearest
neighbors, or they heal (become inactive) spontaneously. More rigorously, the contact process
is a continuous-time Markov process during which active sites become inactive at a rate $\mu$,
while inactive sites turn active at a rate $\lambda m/(2d)$ where $m$ is the number of active
nearest neighbor sites.
The healing rate $\mu$ and the infection rate $\lambda$ are external parameters.

The long-time state of the contact process is determined by the ratio of these two rates.
If $\mu \gg \lambda$, healing occurs much more often than infection. Thus, all infected sites
will eventually become inactive, and the absorbing state without any active sites is the
only steady state. Consequently  the system is in the inactive phase for $\mu \gg \lambda$.
In the opposite limit, $\lambda \gg \mu$, the infection survives for infinite times, i.e.,
there is a steady state with a nonzero density of active sites. This is the active phase.
These two phases are separated by a nonequilibrium phase transition in the DP universality class
occurring at some critical value of the ratio $\lambda/\mu$.

Following Hinrichsen \cite{Hinrichsen97}, we now generalize the contact process to $n$ absorbing
states. Each lattice site can now be in one of $n+1$ states, the active
state A or one of  the $n$ different inactive states I$_k$ ($k=1\ldots n$).
$k$ is sometimes referred to as the ``color'' index. The Markov dynamics
of the generalized contact process is defined via the following transition rates for
pairs of nearest-neighbor sites,
\begin{eqnarray}
w(\textrm{AA} \to \textrm{AI}_k) = w(\textrm{AA} \to \textrm{I}_k\textrm{A}) &=& \bar\mu/n~,
\label{eq:rate_barmu}\\
w(\textrm{AI}_k \to \textrm{I}_k\textrm{I}_k) = w(\textrm{I}_k\textrm{A} \to \textrm{I}_k\textrm{I}_k) &=& \mu_k~,\\
w(\textrm{AI}_k \to \textrm{AA}) = w(\textrm{I}_k\textrm{A} \to \textrm{AA}) &=& \lambda~,
\label{eq:rate_lambda}\\
w(\textrm{I}_k\textrm{I}_l \to \textrm{I}_k\textrm{A}) = w(\textrm{I}_k\textrm{I}_l \to \textrm{A}\textrm{I}_l) &=&
\sigma~,
\label{eq:rate_sigma}
\end{eqnarray}
with $k,l=1\ldots n$ and $k \ne l$. All other transition rates vanish. We are mostly interested in
the fully symmetric case, $\mu_k \equiv \mu$ for all $k$. For $n=1$ and $\bar \mu = \mu$,
the so defined generalized contact process coincides with the simple contact process
discussed above. One of the rates $\bar \mu, \mu, \lambda$, and $\sigma$ can be set to
unity without loss of generality, thereby fixing the unit of time. We choose $\lambda=1$
in the following. Moreover, to keep the parameter space manageable, we focus on the case
$\bar \mu =\mu$ \footnote{We studied the phase diagram for $\bar \mu \ne \mu$ in one space
dimension in Ref.\ \cite{LeeVojta10}. We found that the qualitative behavior is the same as
in the $\bar \mu = \mu$ case. We expect the same to be true in two space dimensions.}.

The rate (\ref{eq:rate_sigma}) is responsible for the new physics in the generalized contact process.
It prevents inactive domains of different color
(different $k$) to stick together indefinitely. By creating active sites at the domain wall,
the two domains can separate. Thus, the rate (\ref{eq:rate_sigma}) allows the domain walls to move
through space.
We emphasize that without the process (\ref{eq:rate_sigma}), i.e., for $\sigma=0$,
the color of the inactive sites becomes unimportant, and all $\textrm{I}_k$ can be
identified. Consequently, for $\sigma=0$, the dynamics of the generalized contact process reduces
to that of the simple contact process for all values of $n$. In the main part of this paper,
we shall focus on the case of $n=2$ inactive states.

Before we turn to our Monte-Carlo simulations of the two-dimensional generalized contact process,
let us briefly summarize the simulation results in one dimension \cite{LeeVojta10} for comparison.
For $\sigma=0$, i.e., in the absence of the boundary activation process (\ref{eq:rate_sigma}),
the system undergoes an absorbing state transition at a healing rate $\mu=\mu_c^{cp}\approx 0.303$,
which agrees with the critical healing rate of the simple contact process. In agreement with the
general arguments above, this transition is in the DP universality class. For healing rates between
$\mu_c^{cp}$ and $\mu^\ast\approx 0.552$, the system is inactive if $\sigma=0$ but an infinitesimal
nonzero $\sigma$ takes it to the active phase. Finally, for $\mu > \mu^\ast$, the transition occurs
at a finite nonzero value of $\sigma$. The one-dimensional generalized contact process
with two inactive states thus has two
lines of phase transitions, (i) the generic transition occurring at $\mu>\mu^*$
and $\sigma=\sigma_c(\mu)>0$ and (ii) the transition occurring for $\mu_c^{cp}<\mu<\mu^*$
as $\sigma$ approaches zero.

%%%%%%%%%%%%%%%%%%%%%%%%%%%%%%%%%%%%%%%%%%%%%%%%%%%%%%%%%%%%%%%%%%%%%%%%%%%%%%%%%
\section{Monte-Carlo simulations}
\label{sec:simulations}
%%%%%%%%%%%%%%%%%%%%%%%%%%%%%%%%%%%%%%%%%%%%%%%%%%%%%%%%%%%%%%%%%%%%%%%%%%%%%%%%%
%%%%%%%%%%%%%%%%%%%%%%%%%%%%%%%%%%%%%%%%%%%%%%%%%%%%%%%%%%%%%%%%%%%%%%%%%%%%%%%%%
\subsection{Method and phase diagram}
\label{subsec:method}
%%%%%%%%%%%%%%%%%%%%%%%%%%%%%%%%%%%%%%%%%%%%%%%%%%%%%%%%%%%%%%%%%%%%%%%%%%%%%%%%%

In order to address the two main problems raised in the introduction, \emph{viz},
the phase diagram of the two-dimensional generalized contact process with two
inactive states and the critical
behavior of its phase transitions, we performed two types of
large-scale Monte Carlo simulations, (i) decay runs and (ii) spreading runs.
Decay runs start from a completely active lattice; we measure the time evolution of the
density $\rho(t)$ of active sites as well as the densities $\rho_1(t)$ and $\rho_2(t)$ of
sites in inactive states I$_1$ and I$_2$, respectively. Spreading simulations start from
a single active (seed) site embedded in a system of sites in state I$_1$.  Here we
monitor the survival probability $P_s(t)$, the number of sites in the active cloud
$N_s(t)$ and the mean-square radius of this cloud, $R^2(t)$.

In both types of runs, the simulation is a sequence of individual events. In each event,
a pair of nearest-neighbor sites is
randomly selected from the active region. For the spreading simulations, the active
region initially consists of the seed site and its neighbors; it is updated in the course
of the simulation according to the actual size of the active cluster. For the decay runs,
the active region comprises the entire sample. The selected pair than undergoes one of the possible
transitions according to eqs.\ (\ref{eq:rate_barmu}) to (\ref{eq:rate_sigma}) with
probability $\tau w$. Here the time step $\tau$ is a constant which we fix at 1/2.
The time increment associated with the event is $\tau/N_{pair}$ where $N_{pair}$
is the number of nearest-neighbor pairs in the active region.

\begin{figure}[tb]
\includegraphics[width=\columnwidth]{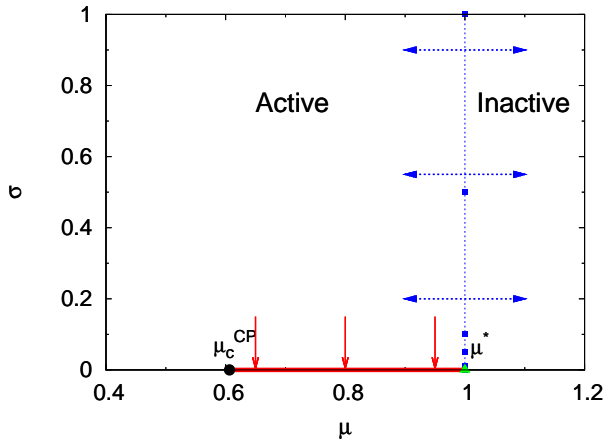}
\caption{(Color online) Phase diagram of the two-dimensional generalized contact process
with two inactive states as function of the healing rate $\mu$ and
the domain-boundary activation rate $\sigma$. For $\mu < \mu_c^{cp} = 0.6066$, the system is
in the active phase for any $\sigma$. For $\mu_c^{cp} < \mu < \mu^* = 1.0000$,
the system is inactive at $\sigma=0$ (thick solid red line), but an infinitesimal
$\sigma$ takes it to the active phase. For $\mu >\mu^\ast$, the system is
inactive for any $\sigma$.}
\label{fig:pd}
\end{figure}

Using this procedure, we investigated the parameter region $0.5 \le \mu \le 1.2$ and  $0\le \sigma \le1$.
We simulated samples with sizes up to $20000 \times 20000$ sites for times up to
$t_{max}= 3\times 10^6$. The $\sigma-\mu$ phase diagram that emerged from these calculations
is shown in in Fig.\ \ref{fig:pd}.

In many respects, it is similar to the phase diagram of the one-dimensional generalized
contact process \cite{LeeVojta10}. In the absence of the domain-boundary activation process
(i.e., for $\sigma=0$), the transition from the active phase to the inactive phase occurs
at a healing rate of $\mu=\mu_c^{cp}=0.6066(2)$ which agrees well with the critical
point of the simple contact process (see, e.g., Refs.\ \cite{Dickman99,VojtaFarquharMast09}).
For healing rates in the interval $\mu_c^{cp} < \mu < \mu^* = 1.0000(2)$,
the generalized contact process is inactive at $\sigma=0$, but an infinitesimal
nonzero $\sigma$ takes it to the active phase. Thus, we find a line a phase transitions
at $\mu_c^{cp} < \mu < \mu^*$ and $\sigma=0$. In addition to this line of $\sigma=0$
absorbing state transitions, we also find a line of generic (nonzero $\sigma$ and $\mu$)
transitions. In contrast to one space dimension, this line is exactly ``vertical''
within our accuracy,
i.e., the critical healing rate $\mu_c= 1.0000(2)$ does \emph{not} depend on $\sigma$ for
all $\sigma>0$. We note in passing that our critical
healing rate is in agreement with the estimate $\mu_c \approx 0.99(1)$ obtained in Ref.\
\cite{Hinrichsen97} for $\sigma=1$.

In the following subsections we shall discuss in detail the properties of both
phase transition lines as well as special point $(\mu^*,0)$
 that separates them.

%%%%%%%%%%%%%%%%%%%%%%%%%%%%%%%%%%%%%%%%%%%%%%%%%%%%%%%%%%%%%%%%%%%%%%%%%%%%%%%%%
\subsection{Generic transition}
\label{subsec:generic}
%%%%%%%%%%%%%%%%%%%%%%%%%%%%%%%%%%%%%%%%%%%%%%%%%%%%%%%%%%%%%%%%%%%%%%%%%%%%%%%%%

\begin{figure}[tb]
\includegraphics[width=\columnwidth]{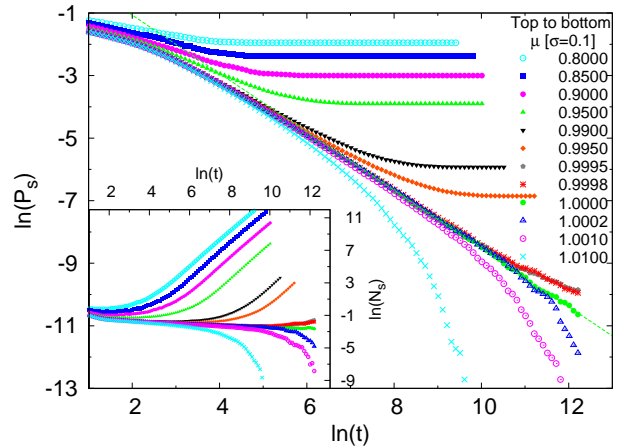}
\caption{(Color online) Spreading simulations at $\sigma=0.1$ for several $\mu$ close to the phase boundary.
Main panel: Survival probability $P_s$
as a function of time $t$. Inset: Number $N_s$ of active sites as a function of time $t$. The data close to criticality are averages over $10^6$ runs on a $4000 \times 4000$
system,
smaller numbers of runs were used away from criticality.}
\label{fig:generic_Ns_Ps}
\end{figure}

In order to identify the generic transition and to study its critical behavior,
we performed sets of spreading simulations at constant domain-boundary activation rate
$\sigma=0.01, 0.05, 0.1, 0.5$ and 1. For each $\sigma$, we have varied
the healing rate $\mu$ from 0.8 to 1.1. Figure \ref{fig:generic_Ns_Ps}
shows the resulting time evolution of the survival probability $P_s$
and the number of sites in the active cloud $N_s(t)$ for $\sigma=0.1$ and several
$\mu$.
The data indicate a critical healing rate of $\mu_c=1.0000(2)$ for this $\sigma$ value.
Analogous simulations for $\sigma=0.01, 0.05, 0.5$ and 1 yielded, somewhat surprisingly,
exactly the same critical healing rate. We thus conclude that in the two-dimensional
generalized contact process, the critical healing rate $\mu_c$ is independent of $\sigma$
for all $\sigma>0$.

\begin{figure}[tb]
\includegraphics[width=\columnwidth]{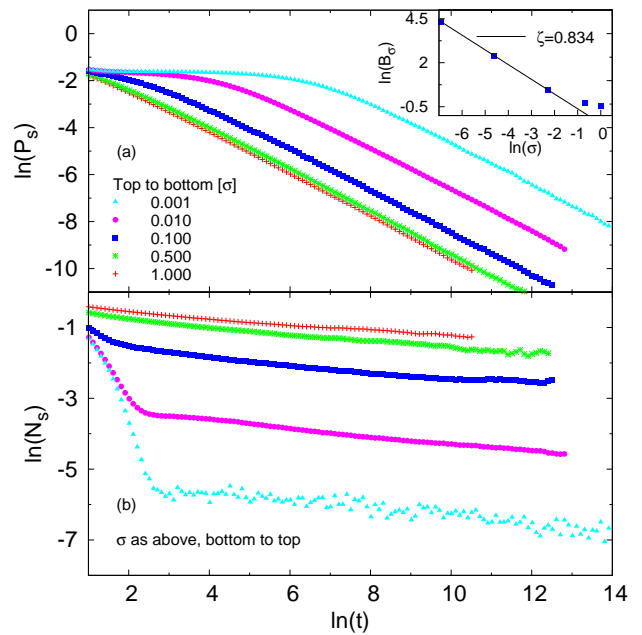}
\caption{(Color online) Survival probability $P_s$ and number of
          active sites $N_s$ as functions of $t$ for several points located on the
          generic phase boundary $\mu=1.0000$ ($2\times 10^6$ to $10^7$ runs used).
          Inset: prefactor $B_\sigma$ vs. $\sigma$. The straight line is a fit
          to a power-law $B_\sigma \sim \sigma^{-\zeta}$.  }
\label{fig:generic_all_critical_Ns_Ps}
\end{figure}

Figure \ref{fig:generic_all_critical_Ns_Ps} shows the survival probability $P_s$ and
number $N_s$ of active sites as functions of time for all the respective critical
points.
In log-log representation, the long-time parts of the $N_s$ and $P_s$ curves for different $\sigma$
are perfectly parallel within their statistical errors, i.e., they differ only by constant factors, confirming that the critical
behavior of the generic transition is universal. Fits of the long-time behavior to the
pure power laws
$P_s =B_\sigma  t^{-\delta}$ and $N_s = C_\sigma t^{\Theta}$ give estimates of
$\delta=0.900(15)$ and $\Theta=-0.100(25)$. These values are very close to the mean-field values
$\delta_{MF}=1$ and $\Theta_{MF}=0$. According to the conjecture by Dornic \emph{et al.}\
\cite{DCCH01}, the generic transition should be in the GV
universality class. Because the upper critical dimension of this universality
class is exactly two, this conjecture corresponds to mean-field behavior with
logarithmic corrections.

\begin{figure}[tb]
\includegraphics[width=\columnwidth]{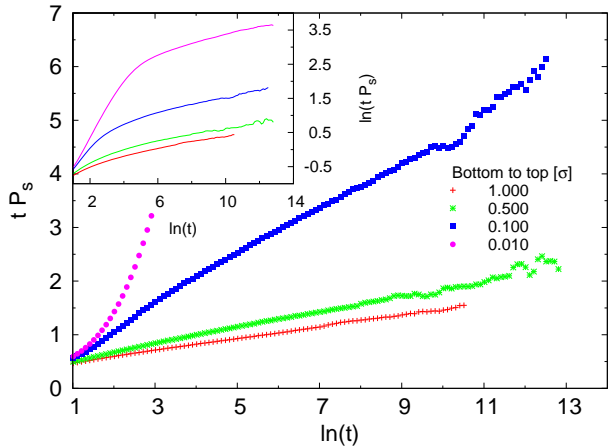}
\caption{(Color online) Survival probability $P_s(t)$ for several points located on the
          generic phase boundary plotted as $t\,P_s $ vs. $\ln(t)$. Straight lines
          correspond to mean-field behavior with logarithmic corrections. Inset: Same data
          plotted as $\ln(t\,P_s)$ vs. $\ln(t)$. Straight lines represent pure power laws.}
\label{fig:GV_comparison}
\end{figure}

To test this prediction we compare in Fig.\ \ref{fig:GV_comparison}
plots of $\ln(t\,P_s)$ vs.\
$\ln(t)$ (straight lines corresponds to power laws) and $t\,P_s$ vs.\
$\ln(t)$ (straight lines
correspond to logarithmic behavior).
Although both functional forms describe the long-time data
reasonably well, the curves in the $\ln(t\,P_s)$ vs.\ $\ln(t)$ plot show a systematic downward
curvature. Moreover, the semi-logarithmic plot, $t\,P_s$ vs.\ $\ln(t)$, leads to straight lines
over a longer
time interval which we take as evidence for GV critical behavior. We performed an
analogous analysis for number of active sites $N_s$. Again, both a simple power law and
mean-field behavior with logarithmic corrections describe the data reasonably well,
with the quality of fits being somewhat higher for the latter case. We also measured
(not shown) the mean-square radius $R^2(t)$ of the active cloud
as a function of time. A pure power-law fit of its long time behavior, $R^2(t) \sim t^{2/z}$,
gives $2/z=0.97(4)$ ($z=2.06(8)$). The data can be described equally well by mean-field behavior
$R^2(t)~\sim t$ with logarithmic corrections.

\begin{figure}[tb]
\includegraphics[width=\columnwidth]{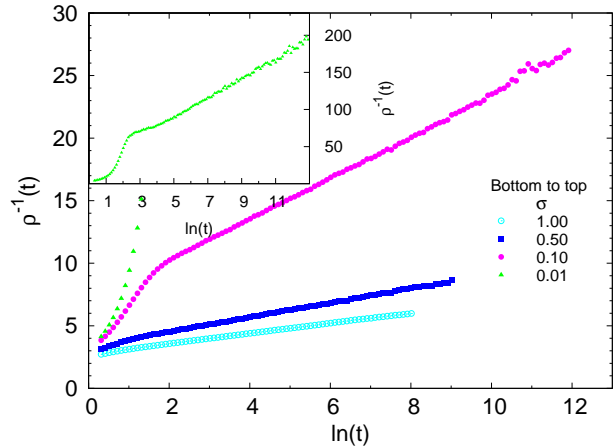}
\caption{(Color online) Density of active sites plotted as $\rho^{-1}(t)$ vs.\ $\ln(t)$
for several points located on the generic phase boundary. The data are averages over
100 runs with system size 500 $\times$ 500. The curve for $\sigma=0.01$ is shown in the inset
because its density values are much smaller than those of the other curves.
}
\label{fig:generic_rho}
\end{figure}

In addition to the spreading runs, we also performed density decay runs at the generic
phase boundary. The resulting density of active sites $\rho$ as a function of time can
be fitted with a pure power law $\rho(t) \sim t^{-\alpha}$ giving a very small
value of $\alpha = 0.080(4)$. A better fit is achieved with the simple logarithmic time dependence
$\rho(t) \sim 1/\ln(t/t_0)$ (with $t_0$ a microscopic time scale) expected for the GV
universality class. This type of behavior is demonstrated in
Fig.\ \ref{fig:generic_rho}.

In summary, although all our results for the generic transition can be fitted
both by pure power laws and by mean-field behavior with logarithmic corrections,
the latter functional forms yield fits of somewhat higher quality. We also note
that the critical exponents resulting from the pure power-law fits approximately
fulfill the hyperscaling relation $\Theta -d/z = -\alpha-\delta$. However, the
agreement is not very good (in particular, it is significantly worse than in
one dimension \cite{LeeVojta10}), indicating that the measured pure power-laws
are not the true asymptotic behavior.
Our results thus support
the conjecture that the generic transition of the two-dimensional generalized
contact process with two inactive states is in the GV universality class.

%%%%%%%%%%%%%%%%%%%%%%%%%%%%%%%%%%%%%%%%%%%%%%%%%%%%%%%%%%%%%%%%%%%%%%%%%%%%%%%%%
\subsection{Transition at $\sigma=0$}
\label{subsec:sigma0}
%%%%%%%%%%%%%%%%%%%%%%%%%%%%%%%%%%%%%%%%%%%%%%%%%%%%%%%%%%%%%%%%%%%%%%%%%%%%%%%%%

After addressing the generic transition, we now discuss in more detail the line of
phase transitions occurring at $\sigma=0$ and $\mu_c^{cp}<\mu<\mu^\ast$. To study these transitions,
we carried out several sets of simulations for fixed healing rate
$\mu$ and several $\sigma$ values approaching $\sigma=0$.

\begin{figure}
\includegraphics[width=\columnwidth]{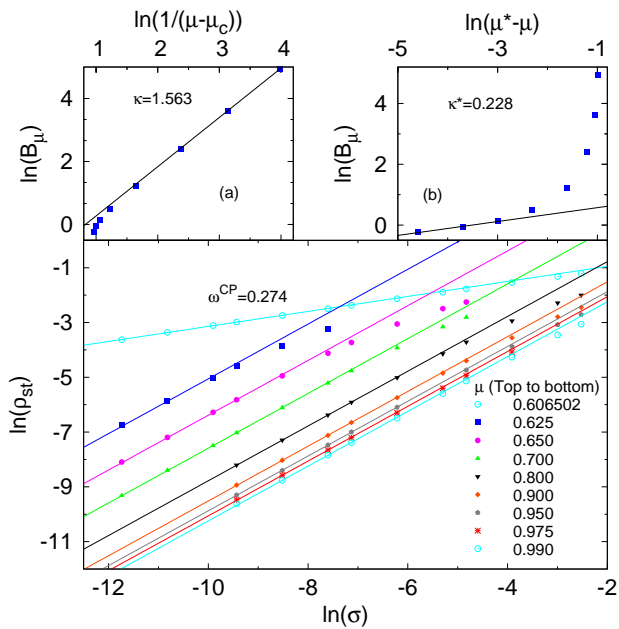}
\caption{(Color online) Density decay simulations. Main panel: stationary density $\rho_{st}$ as a function
of the boundary rate $\sigma$ for various healing rates $\mu$. For $\mu_c^{cp} < \mu <\mu^*$,
the solid lines are fits of the low-$\sigma$ behavior to
$\rho_{st} = B_\mu \sigma$. At the simple contact process critical point, $\mu=\mu_c^{cp}=0.6066$,
we fit to the power-law $\rho_{st} \sim \sigma^{\omega_{cp}}$ which gives an exponent of $\omega_{cp} = 0.274(5)$.
The data are averages over 300 to 600 runs with system sizes $100 \times 100$.
Inset a: prefactor $B_\mu$ of the linear $\sigma$
dependence
as a function of $\mu-\mu_c^{cp}$. A fit to a power law gives $B_\mu \sim (\mu-\mu_c^{cp})^{-\kappa}$
with $\kappa= 1.56(5)$. Inset b: prefactor $B_\mu$
as a function of $\mu^*-\mu$. A fit to a power law gives $B_\mu \sim (\mu^*-\mu)^{\kappa^*}$
with $\kappa^*\approx0.23$.}
\label{fig:rho_st}
\end{figure}

We start by discussing the density decay runs. Figure \ref{fig:rho_st}
shows the stationary density $\rho_{st}$ of active sites (reached at long times)
as function of $\sigma$
for several values of the healing rate $\mu$.
The figure shows that the stationary density depends linearly on $\sigma$ for all healing rates
in the interval $\mu_c^{cp} < \mu < \mu^\ast$, i.e., $\rho_{st} = B_\mu
\sigma^\omega$ with $\omega=1$ and $B_\mu$ being a $\mu$-dependent constant.
We also analyzed how the prefactor $B_\mu$ of this mean-field-like  behavior depends on
the distances from the simple contact process critical point and from the
special point at $\mu=\mu^\ast$ and $\sigma=0$. As inset (a) of
Fig.\ \ref{fig:rho_st} shows, $B_\mu$ diverges as $(\mu-\mu_c^{cp})^{-\kappa}$ with
$\kappa= 1.56(5)$. According to inset (b), it vanishes as $(\mu^\ast-\mu)^{\kappa^*}$
with $\kappa^\ast \approx 0.23$ when approaching $\mu^\ast$.

At the critical healing rate $\mu_c^{cp}$ of the simple contact
process, the stationary density displays a weaker $\sigma$-dependence. A fit to a
power-law $\rho_{st} \sim \sigma^{\omega_{cp}}$ gives an exponent value of
$\omega_{cp}= 0.274(5)$.

\begin{figure}
\includegraphics[width=\columnwidth]{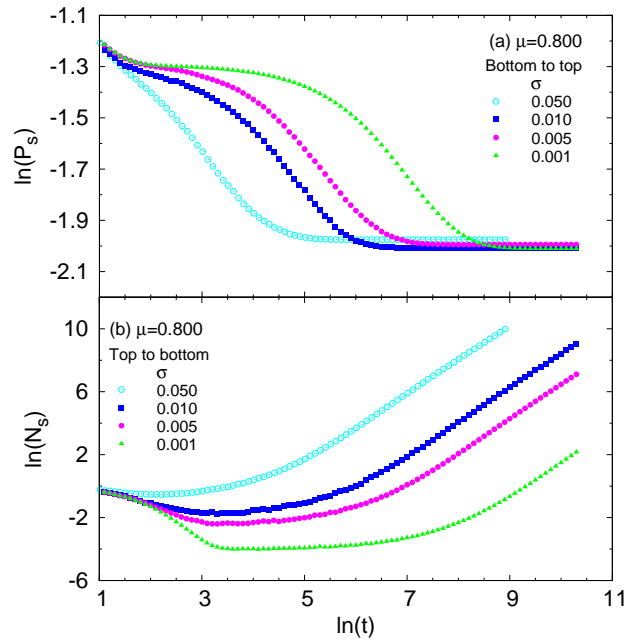}
\caption{(Color online) Spreading simulations: Survival probability $P_s$ and number of active sites $N_s$
as functions of time $t$ for a fixed healing rate of $\mu=0.8$ and several $\sigma$.
The data are averages over 2000 to 10000 runs on a $4000 \times 4000$ system.
}
\label{fig:sigma0_Ns_Ps}
\end{figure}

Let us now compare these results with the behavior of spreading simulations
in the same parameter region. Figure \ref{fig:sigma0_Ns_Ps} shows the survival probability
$P_s(t)$ and the number of active sites $N_s(t)$ for a fixed healing rate of $\mu=0.8$
and several values of the boundary rate $\sigma$.
After an initial decay, the number of active sites grows with time for all $\sigma$ values,
establishing that the system is in the active phase for all $\sigma>0$. In agreement with this,
the survival probability approaches a nonzero constant in the long-time limit.
Remarkably, this stationary survival probability does \emph{not} approach zero with vanishing $\sigma$.
Instead, it approaches a $\sigma$-independent constant. We performed similar sets of
simulations at other values of $\mu$ in the range $\mu_c^{cp} < \mu < \mu^*$, with
analogous results.

We thus conclude that the behavior at the $\sigma=0$ transition of the
two-dimensional generalized contact process is very similar to the one-dimensional
case. It can be understood in terms of the domain-wall motion as follows \cite{LeeVojta10}.
The relevant long-time degrees of freedom at $\mu > \mu_c^{cp}$ and $\sigma \ll 1$ are
the domain walls between I$_1$ and I$_2$ domains.
These walls can hop, branch and annihilate. The crucial observation is that the
rates which control the domain wall dynamics are all proportional to
$\sigma$ for $\sigma \ll 1$, implying that their ratios are $\sigma$-independent.
Consequently, the stationary state of the domain walls does not depend on $\sigma$
for $\sigma \ll 1$.  This explains why the survival probability $P_s$ saturates
at a nonzero, $\sigma$-independent value in Fig.\ \ref{fig:sigma0_Ns_Ps}.
It also explains the $\sigma$-dependence of the stationary density $\rho_{st}$
because active sites are created mostly at the domain walls at rate $\sigma$. Therefore,
their stationary density is proportional to both $\sigma$ and the stationary domain
wall density $\rho_{dw}$, i.e., $\rho_{st} \sim \sigma \rho_{dw}$, in agreement with
Fig.\ \ref{fig:rho_st}. Based on this argument, the exponent $\kappa^\ast$ in inset (b) of Fig.\
\ref{fig:rho_st} should be identical to the exponent $\beta$
of the generic transition line \cite{LeeVojta10}, which vanishes in mean-field theory.
Our value, $\kappa^\ast\approx 0.23$ is thus somewhat too high which we attribute to
it not representing the asymptotic behavior, in agreement with the significant
curvature of the data in inset (b) of Fig.\ \ref{fig:rho_st}.

Just as in one dimension, the phase transition line at $\sigma=0$ and
$\mu_c^{cp} <\mu <\mu^\ast$ is thus not a true critical line. It only appears
critical because the stationary density $\rho_{st}$ (trivially) vanishes with
$\sigma$. Correspondingly, the time evolution right on the transition line
$\sigma=0$ does not display critical power laws.
This also implies that the point $(\mu,\sigma)=(\mu^\ast,0)$ is not a multicritical
point, but a simple critical point in the same universality class as the generic
transition.

%%%%%%%%%%%%%%%%%%%%%%%%%%%%%%%%%%%%%%%%%%%%%%%%%%%%%%
\subsection{Scaling of $\rho_{st}$ at the contact process critical point $(\mu_c^{cp},0)$}
\label{subsec:CP_contact}
%%%%%%%%%%%%%%%%%%%%%%%%%%%%%%%%%%%%%%%%%%%%%%%%%%%%%%

The behavior of the stationary density of active sites $\rho_{st}$ close to the
simple contact process critical point at $\mu=\mu_c^{cp}$ and $\sigma=0$
can be understood in terms of a phenomenological scaling theory.
We assume  the homogeneity relation
\begin{equation}
\rho_{st}(\Delta \mu, \sigma) = b^{\beta_{cp}/\nu^{\perp}_{cp}} \rho_{st} (\Delta \mu\, b^{-1/\nu^{\perp}_{cp}}, \sigma b^{-y_{cp}})
\label{eq:rho_homogeneity}
\end{equation}
where $\Delta \mu = \mu - \mu_c^{cp}$, and $b$ is an arbitrary scale factor.
$\beta_{cp}=0.584$ and $\nu^\perp_{cp}=0.734$ are the usual order parameter and correlation
length exponents of the two-dimensional contact process \cite{Dickman99,VojtaFarquharMast09},
and $y_{cp}$ denotes the scale dimension of $\sigma$ at this
critical point. Setting $b=\sigma^{1/y_{cp}}$ gives rise to the scaling form
\begin{equation}
\rho_{st}(\Delta \mu, \sigma) = \sigma^{\beta_{cp}/(\nu^\perp_{cp}y_{cp} )} X\left(
\Delta \mu\, \sigma^{-1/(\nu^\perp_{cp}y_{cp} )}\right)~
\label{eq:rho_scaling_cp}
\end{equation}
where $X$ is a scaling function.
At criticality, $\Delta\mu =0$, this leads to $\rho_{st}(0,\sigma) \sim \sigma^{\beta_{cp}/(\nu^\perp_{cp}y_{cp}
)}$ (using $X(0)=\textrm{const}$). Thus, $\omega_{cp}=
\beta_{cp}/(\nu^\perp_{cp}y_{cp})$. For $\sigma \to 0$ at nonzero $\Delta
\mu$, we need the large-argument limit of the scaling function $X$.
On the active side of the critical point, $\Delta \mu <0$, the scaling function
behaves as $X(x) \sim |x|^{\beta_{cp}}$ to reproduce the correct critical behavior
of the density, $\rho_{st} \sim |\mu-\mu_c^{cp}|^{\beta_{cp}}$.

On the inactive side of the critical point, i.e., for $\Delta \mu >0$ and $\sigma \to 0$,
we assume the scaling function to behave as
$X(x) \sim x^{-\kappa}$. We thus obtain
$\rho_{st} \sim (\Delta \mu)^{-\kappa} \sigma^\omega$
(just as observed in Fig.\ \ref{fig:rho_st}) with $\omega=(\beta_{cp}+\kappa)/(\nu^\perp_{cp}y_{cp})$.
As a result of our scaling theory, the exponents $\omega, \omega_{cp}$ and $\kappa$ are not independent, they need to
fulfill the relation $\omega_{cp} (\beta_{cp} +\kappa) = \beta_{cp} \omega$.
Our numerical values, $\omega=1$, $\omega_{cp}= 0.274$ and $\kappa= 1.56$ fulfill this relation
in very good approximation, indicating that
they represent asymptotic exponents and validating the homogeneity relation
(\ref{eq:rho_homogeneity}). The resulting value for the scale dimension $y_{cp}$ of
$\sigma$ at the simple contact process critical point is $y_{cp}=2.9(1)$.

%%%%%%%%%%%%%%%%%%%%%%%%%%%%%%%%%%%%%%%%%%%%%%%%%%%%%%%%%%%%%%%%%%%%%%%%%%%%%%%%%
\section{Conclusions}
\label{sec:conclusions}
%%%%%%%%%%%%%%%%%%%%%%%%%%%%%%%%%%%%%%%%%%%%%%%%%%%%%%%%%%%%%%%%%%%%%%%%%%%%%%%%%

To summarize, we investigated the two-dimensional generalized contact process with two inactive
states by means of large-scale Monte-Carlo simulations. Its global phase diagram is very similar
to that of the corresponding one-dimensional model. In particular, the generic ($\sigma>0$)
phase boundary between the active and inactive phases does not continuously connect to the
critical point of the $\sigma=0$ problem, i.e., the critical point $(\mu_c^{cp},0)$ of the simple contact process.
Instead, it terminates at a separate end point ($\mu^\ast,0)$ on the $\mu$ axis.
As a result, the two-dimensional generalized contact process has two nonequilibrium phase transitions.
In addition to the generic transition occurring for $\sigma>0$, there is a line of transitions
at $\sigma=0$ and $\mu_c^{cp} < \mu < \mu^\ast$. We note that there is one interesting
difference between the phase diagrams in one and two dimensions. In one dimension, the critical healing rate $\mu_c$
increases with increasing boundary rate $\sigma$. In contrast, the results of this paper show
that the critical healing rate in two dimensions is completely independent of $\sigma$.
Moreover, its value seems to be equal to unity (i.e., equal to that of the infection rate $\lambda$).
The reason for this
peculiar behavior is presently an open question.

To determine the critical behavior of the generic transition, we performed simulations
at and close to several points on the generic ($\sigma>0$) phase boundary. We found the
same critical behavior for all of these points, i.e, it is universal. Our data can be fitted
reasonably well with pure power laws, giving the exponents $\Theta=-0.100(25)$, $\delta=0.900(15)$, $\alpha=0.080(4)$,
and $z=2.06(8)$. However, fits of equal and sometimes even better quality over longer ranges of time can be obtained
by fitting to mean-field critical behavior, $\Theta=0$, $\delta=1$, $\alpha=0$, and $z=2$
with logarithmic corrections. Our results thus support the conjecture
\cite{DCCH01} that the critical behavior of the two-dimensional generalized contact
process is right at its upper critical dimensions. (This implies that the DP2 class
in two dimensions coincides with the GV class.).
We also note that our simulations showed no indications of the transition being split
into a symmetry-breaking transition and a separate DP transition as found in
some absorbing-state Potts models \cite{DrozFerreiraLipowski03}.

As in one space dimension, the line of transitions
at $\sigma=0$ and $\mu_c^{cp} < \mu < \mu^\ast$ is not a critical line. The survival probability
$P_s$ remains finite when approaching this line. The density $\rho$ of active sites
vanishes, but simply because the domain-boundary activation rate $\sigma$ vanishes. The behavior in
the vicinity of the transition line is controlled by the dynamics of the I$_1$-I$_2$ domain
walls which is not critical for $\mu_c^{cp} < \mu < \mu^\ast$.

Crossovers between various universality classes of absorbing state transitions in \emph{one
dimension} have been investigated by several authors \cite{OdorMenyhard08,ParkPark07,ParkPark08,ParkPark09}.
Some of the scenarios lead to conventional crossover scaling (of the type
$\sigma_c \sim (\mu - \mu_c^{cp})^{1/\phi}$). Park and Park \cite{ParkPark07} found a
discontinuous jump in the phase boundary along the so-called excitatory route from
infinitely many absorbing states to a single absorbing state. There also is some similarity
between our mechanism and the so-called channel route \cite{ParkPark08} from the PC universality
class to the DP class which involves an infinite number of absorbing states characterized
by an auxiliary density (which is density of I$_1$-I$_2$ domain walls in
one-dimensional the generalized contact process \cite{LeeVojta10}). To the best of our
knowledge, a similarly systematic investigation of crossovers between absorbing
state universality classes in two space dimensions has not yet been performed.

As our results suggest that the two-dimensional generalized contact process is right
at the upper critical dimensions, the critical behavior of its (generic) phase transition in
dimensions $d>2$ should be governed by mean-field theory.

%%%%%%%%%%%%%%%%%%%%%%%%%%%%%%%%%%%%%%%%%%%%%%%%%%%%%%%%%%%%%%%%%%%%%%%%%%%%%%%%%
\section*{Acknowledgements}
%%%%%%%%%%%%%%%%%%%%%%%%%%%%%%%%%%%%%%%%%%%%%%%%%%%%%%%%%%%%%%%%%%%%%%%%%%%%%%%%%

We acknowledge helpful discussions with Ronald Dickman, Geza Odor and Hyunggyu Park.
This work has been
supported in part by the NSF under grant no. DMR-0339147 and DMR-0906566 as well as by
Research Corporation.

\bibliographystyle{apsrev4-1}
\bibliography{../../00Bibtex/rareregions}
\end{document}